# A Generalized Genetic Random Field Method for the Genetic Association Analysis of Sequencing Data

**Ming Li**[1,†], **Zihuai He**[2,†], **Min Zhang**[2], **Xiaowei Zhan**[2], **Changshuai Wei**[3], **Robert C. Elston**[4], and **Qing Lu**[3,*]

[1]Division of Biostatistics, Department of Pediatrics, University of Arkansas for Medical Sciences, Little Rock, Arkansas, United States of America

[2]Department of Biostatistics, University of Michigan, Ann Arbor, Michigan, United States of America

[3]Department of Epidemiology and Biostatics, Michigan State University, East Lansing, Michigan, United States of America

[4]Department of Epidemiology and Biostatistics, Case Western Reserve University, Cleveland, Ohio, United States of America

## Abstract

With the advance of high-throughput sequencing technologies, it has become feasible to investigate the influence of the entire spectrum of sequencing variations on complex human diseases. Although association studies utilizing the new sequencing technologies hold great promise to unravel novel genetic variants, especially rare genetic variants that contribute to human diseases, the statistical analysis of high-dimensional sequencing data remains a challenge. Advanced analytical methods are in great need to facilitate high-dimensional sequencing data analyses. In this article, we propose a generalized genetic random field (GGRF) method for association analyses of sequencing data. Like other similarity-based methods (e.g., SIMreg and SKAT), the new method has the advantages of avoiding the need to specify thresholds for rare variants and allowing for testing multiple variants acting in different directions and magnitude of effects. The method is built on the generalized estimating equation framework and thus accommodates a variety of disease phenotypes (e.g., quantitative and binary phenotypes). Moreover, it has a nice asymptotic property, and can be applied to small-scale sequencing data without need for small-sample adjustment. Through simulations, we demonstrate that the proposed GGRF attains an improved or comparable power over a commonly used method, SKAT, under various disease scenarios, especially when rare variants play a significant role in disease etiology. We further illustrate GGRF with an application to a real dataset from the Dallas Heart Study. By using GGRF, we were able to detect the association of two candidate genes, *ANGPTL*3 and *ANGPTL*4, with serum triglyceride.

[*]Correspondence to: Qing Lu, Department of Epidemiology and Biostatics, Michigan State University, East Lansing, MI 48824, United States of America. qlu@epi.msu.edu.
[†]These authors contributed equally to this article.

**Keywords**

rare variants; small-scale sequencing studies; generalized estimating equation

## Introduction

Genome-wide association studies (GWAS) have been commonly used to evaluate the association of millions of single nucleotide polymorphisms (SNPs) with complex human diseases. To date, more than 1,000 diseases-related genetic variants have been revealed by GWAS [Hindorff et al., 2009]. Despite such successes, for most complex diseases the identified genetic variants only account for a small percentage of the disease heritability [Manolio et al., 2009]. One possible explanation is that GWAS are based on the "Common Disease–Common Variant" hypothesis. Under this hypothesis, complex diseases are likely caused by multiple common variants with appreciable frequencies (e.g., >1%), each conferring a small or moderate effect [Schork et al., 2009]. However, evidence from previous studies (e.g., genetic studies of inherited hearing loss and lipid metabolism) suggests that the genetic etiology of complex diseases can be highly heterogeneous [McClellan and King, 2010], and rare variants within a gene or across different genes may collectively have a significant influence on diseases [Bodmer and Bonilla, 2008; Schork et al., 2009]. The fast development of next generation sequencing technologies facilitates the detection of millions of rare sequence variations [Ansorge, 2009; Bodmer and Bonilla, 2008; Schuster, 2008], and enables us to comprehensively assess the potential contribution of rare variants in complex diseases [Eichler et al., 2010]. Meanwhile, the emergence of a large amount of sequence variation also poses a great challenge for the statistical analyses of high-dimensional sequencing data. Advanced statistical methods are in great need to evaluate the role of rare variants, in conjunction with common variants, in complex human diseases.

Standard statistical methods generally have low power for testing individual rare variants because of the low allele frequencies of rare variants. Therefore, a number of methods were proposed to collapse multiple rare variants in a gene, or a genetic region, into one group. Among these methods, the combined multivariate and collapsing method collapses rare variants by evaluating whether any rare allele occurs at any loci for a subject [Li and Leal, 2008]; Morris and Zeggini [2010] suggested collapsing the rare variants by counting the rare alleles across all loci; the weighted sum test calculates a weighted average of rare alleles [Madsen and Browning, 2009], where the weights are based on the minor allele frequency of each rare variant; Han and Pan used a data-adaptive sum test to account for the possible bidirection of genetic effects [Han and Pan, 2010]. These methods, commonly referred to as burden tests, can improve power by combing multiple rare variants into a "super" variant, and thus reduce the burden of multiple testing. A major disadvantage of burden tests is that a minor allele frequency threshold (e.g., <1%) needs to be specified to collapse the variants, which is often arbitrary. Several methods with a data-adaptive threshold were also proposed [Price et al., 2010; Zawistowski et al., 2010]. Nevertheless, these methods usually require a permutation test, which can be computationally intensive.

Unlike burden tests, which form statistics on the collapsed rare variants, similarity-based methods build the statistics by linking the genetic similarity of individuals to their phenotypic similarity. Wessel and Schork discussed a number of choices to calculate genetic similarity, and proposed a multivariate distance matrix regression by using the genetic similarity matrix as response variables [Wessel and Schork, 2006]. Tzeng et al. measured haplotype similarity by the average allelic sharing across variants, and evaluated the gene-trait association by testing the regression coefficient of haplotype similarity on trait-similarity [Tzeng et al., 2009]. This proposed method, referred to as SIMreg, was later adapted as a random-effect model to study gene–environment interactions [Tzeng et al., 2011]. More recently, kernel machine-based methods, such as the sequence kernel association test (SKAT) [Lee et al., 2012; Wu et al., 2010, 2011], have gained popularity. SKAT aggregates the genetic association through a kernel matrix, which is flexible for testing a large number of variants while adjusting for covariates. These similarity-based methods generally avoid selection of thresholds, allow for multiple variants with different directions and magnitudes of effects, and are computationally efficient without requiring a permutation test.

The similarity-based methods build on the concept that, if there is a gene-phenotype association, the genetic similarity between two individuals will lead to their phenotypic similarity. A similar concept has been used in imaging analysis and spatial analysis, in which subjects who are close to each other on a map or in space tend to have similar outcomes. The outcomes of all subjects, each corresponding to a location in a space, form a stochastic process, referred to as a random field with random coordinate variables that are spatially correlated in certain ways [Besag, 1974]. The analogy with spatial statistics motivates the use of a random field to genetic research. A genetic random field model was recently proposed for quantitative phenotypes [He et al., 2013]. In this article, we propose a generalized genetic random field (GGRF) method for the statistical analysis of sequencing data. In this method, we view phenotypes of individuals as a random field in a Euclidean space spanned by their sequenced genotypes. Each individual has a location in this space determined by his/her genotypes. In the presence of an association, individuals tend to have similar phenotypes if they are "adjacent" in the Euclidean space. The proposed method can be applied to phenotypes with a variety of distributions (e.g., quantitative and binary phenotypes), through a GEE framework. It can also integrate into the analysis of both common and rare variants, and evaluate their combined contribution to disease phenotypes. Similar to the existing similarity-based methods (e.g., SIMreg and SKAT), GGRF has a number of advantages. It avoids selection of thresholds, allows for multiple variants with different directions and magnitude of effects, and is computationally efficient for high-dimensional sequencing data analysis. Furthermore, it has a nice asymptotic property, and can be applied to small-scale sequencing data without need of small-sample adjustment. We compare the performance of GGRF with that of SKAT via simulation studies. The proposed method is further illustrated with an application to a sequencing dataset from the Dallas Heart Study (DHS).

## Materials and Methods

### Generalized Genetic Random Field

The GGRF method is motivated by the general idea that, if the genetic variants are jointly associated with the phenotypes, the genetic similarity between subjects contributes to their phenotypic similarity. In GGRF, we map individuals into a Euclidean space, where each individual has a location determined by his/her sequenced genotype. If there is a genotype–phenotype association, we expect individuals who are adjacent in the Euclidean space have more similar phenotypes than those that are further apart. Based on this concept, we model the conditional phenotypic mean of each individual as a linear function of a weighted sum of phenotypes of the remaining individuals, where the weights are determined by the genetic similarity of the individuals [He et al., 2013].

Assume $K$ variants in a gene or a genetic region were sequenced and $M$ covariates (e.g., age) were measured for $N$ subjects. Let $y_i$ be the phenotypic value for the $i$-th subject, $G_i = (g_{i,1}, g_{i,2}, \ldots, g_{i,K})'$ be the genotypes of $K$ sequence variants, and $X_i = (x_{i,1}, x_{i,2}, \ldots, x_{i,M})'$ be the covariates. We can express a phenotypic mean for each individual as a linear function of covariates and a weighted sum of the phenotypes of all other individuals,

$$E\left(y_i | y_{-i}\right) = \mu_i + \gamma \sum_{j \neq i} s_{i,j}\left(y_j - \mu_j\right), \quad (1)$$

where $y_{-i}$ denotes phenotypes for all subjects other than subject $i$ and $\mu_i = f(X_i'\beta)$, where $f(\cdot)$ is the mean function as in a generalized linear model, used for controlling covariates. Specifically, if the phenotype is quantitative, we use the identity link $f(x) = x$; if the phenotype is binary, we use the logistic link $f(x) = \frac{\exp(x)}{1 + \exp(x)}$. $s_{i,j}$ is a weight representing the relative contribution of the $j$-th subject in predicting the phenotype of subject $i$, which is determined by the genetic similarity between subjects $i$ and $j$. $\gamma$ is a nonnegative coefficient measuring the ability of all the remaining subjects to predict the phenotype of subject $i$, which can also be interpreted as the magnitude of the joint association of $K$ sequence variants with the phenotype. If none of the sequence variants are associated with the phenotype (i.e., $\gamma = 0$), the phenotype of subject $i$ will be independent of the phenotypes of others, regardless of their genetic similarity. On the other hand, a large $\gamma$ indicates a strong genetic contribution to the phenotype. Therefore, to examine the joint association of $K$ sequence variants with the phenotype, we test a null hypothesis with a single parameter, $H_0 : \gamma = 0$.

### Statistical Inference

In this section, we propose a generalized estimating equation (GEE)-based statistic to test the null hypothesis, $H_0 : \gamma = 0$. For convenience, we rewrite equation (1) in matrix form,

$$E\left(Y | Y_-\right) = f\left(X\beta\right) + \gamma S[Y - f\left(X\beta\right)], \quad (2)$$

where $Y = (y_1, y_2, \ldots, y_N)'$, $X = (X_1, X_2, \ldots, X_N)'$ and $Y_- = (y_{-1}, y_{-2}, \ldots, y_{-N})'$. $\mu = f(X\beta)$, where $\mu = (\mu_1, \ldots, \mu_N)'$ represents the nongenetic mean. $S$ is an $N \times N$ similarity matrix with $s_{i,j}$ as its element in row $i$ and column $j$, $1 \leq i \leq j \leq N$, and zeros on the diagonal. The parameters in equation (2) can be estimated by solving the following unbiased estimating equation:

$$U_\gamma(\beta, \gamma) = \left[ \frac{\partial E(Y|Y_-)}{\partial \gamma} \right]' [Y - E(Y|Y_-)]$$
$$= (Y - \mu)' S(I - \gamma S)(Y - \mu) = 0. \qquad (3)$$

Based on equation (3), we estimate $\gamma$,

$$\hat{\gamma} = \frac{(Y - \hat{\mu})' S(Y - \hat{\mu})}{(Y - \hat{\mu}) S^2 (Y - \hat{\mu})},$$

where the nongenetic mean, $\hat{\mu}$, can be estimated under the null hypothesis $\gamma = 0$. We define $W$ as a diagonal matrix with its $i$-th element $w_i = 1$ for quantitative phenotypes, and $w_i = \hat{\mu}_i(1 - \hat{\mu}_i)$ for binary phenotypes with a logistic link. Large values of $\gamma$ suggest an association of the sequence variants with the phenotype. Given the observed value $\hat{\gamma}$, $\hat{\gamma}_{obs}$, the $P$-value of the association test can be calculated by

$$P_{H_0}(\hat{\gamma} > \hat{\gamma}_{obs}) = P\left((Y - \hat{\mu})'(S - \hat{\gamma}_{obs} S^2)(Y - \hat{\mu}) > 0\right).$$

We note that $(Y - \hat{\mu})'(S - \hat{\gamma}_{obs} S^2)(Y - \hat{\mu})$ asymptotically follows a mixture of Chi-squares, $\sum_{k=1}^{K} \lambda_k \chi_{1,k}^2$, where $(\lambda_1 \ldots, \lambda_K)$ are the eigenvalues of the matrix $P^{1/2}(S - \hat{\gamma}_{obs} S^2)P^{1/2}$ and $P = W - WX(X' WX)^{-1} X' W$ [Wu et al., 2011]. Given the asymptotic distribution, Davies' method can then be used to obtain the significance level of the association test [Davies, 1980].

The test statistic used in GGRF holds a nice asymptotic property for small sample size studies. For quantitative phenotypes, the test statistic $\hat{\gamma}$ is ancillary to the variance of $Y_i$ because the variance term in the numerator and denominator are cancelled out. Without using any asymptotic approximation, the test is an exact test and is therefore not conservative. For binary phenotypes, the estimated variance of $Y$ depends on estimated means. When there is no covariate or covariates only have small or moderate effect on the mean, the test statistic is still not conservative because the estimated variance in the numerator and denominator are also cancelled out or nearly cancelled out. The asymptotic approximation is only needed when the covariates have large impact on the mean.

## Weight and Similarity Functions

Sequencing data comprise a large number of common and rare variants. Although rare variants have low allele frequencies, they could contribute significantly to the phenotype. A good choice of weights and similarity metrics that reflect the contribution of rare variants and the underlying genetic similarity between individuals can improve the power of the association test. In this paper, we consider four commonly used weights. As we discuss later, each set of weights assumes different contributions of rare variants to the disease. Given the prespecified vector of weights, $\omega = (\omega_1, \omega_2, \ldots, \omega_K)'$ for $K$ sequence variants, we propose a general $p$-norm distance-based genetic similarity (NDS) between subject $i$ and subject $j$ as:

$$s_{i,j} = B - \|G_i - G_j\|_p = 2\left(\sum_{k=1}^{K} \omega_k\right)^{1/p} - \left(\sum_{k=1}^{K} \omega_k |g_{i,k} - g_{j,k}|^p\right)^{1/p}, \tag{4}$$

where $\|G_i - G_j\|_p$ is the $p$-norm distance between subjects $i$ and $j$ based on their genotypes, and $B$ is the corresponding mathematical supremum over the distance between any two subjects. Note that the $1^{\text{st}}$ order NDS ($p = 1$) is equivalent to the commonly used identity-by-state (IBS) metric, and the $2^{\text{nd}}$ order NDS ($p = 2$) is based on the commonly used Euclidean distance.

## Results

### Simulation Studies

We conducted simulation studies to compare the performance of GGRF with a commonly used method, SKAT. In the simulations, we varied the underlying disease model, choice of weights, causal variants/noise variants ratios, and similarity metrics. In each case, we compared power and type I error of the two methods. To mimic a real data scenario, the genotype data used in the simulations was based on the exome sequencing data of 697 subjects from the 1000 Genome project [Almasy et al., 2011]. The genotype data comprised 508 sequence variants located on chromosome 22 with minor allele frequencies (MAFs) ranging from 0.07% to 49.4%. The distribution of the MAFs was given in Figure 1. The majority of the 508 sequence variants were rare variants with MAFs less than 1%. The phenotypic values of the samples were simulated based on the genotypes and assumed disease models that were discussed in detail below. In each simulation, type I error and power of the two methods were estimated based on 1,000 replicates. For comparison purposes, IBS was used as the similarity metric for GGRF and SKAT, unless specified otherwise. In the simulation studies described below, we focused on the comparison of GGRF and SKAT for one-direction of effect sizes. The performance of SKAT and other commonly used Burden tests were extensively compared in previous studies [Lee et al., 2012; Wu et al., 2011]. In the Appendix, we also showed Burden test tended to have the highest power under one-direction of effect sizes, but the proposed GGRF with an appropriate kernel was able to share similar advantages with SKAT over Burden test, for being robust to bidirection of effect sizes (See Fig. B1 in Appendix). Such results were consistent with previous studies and were not detailed here.

### Simulations I: Various Prespecified Weights under Various Disease Models—
Both GGRF and SKAT can use prespecified weights to boost the power of association tests, and their performance may depend on how well the weights reflect the relative contribution of the genetic variants to the disease. Without any prior knowledge of the underlying disease model, the weights are often prespecified as a function of MAFs. In this simulation, we chose four weight functions and evaluated their influence on the methods' performance.

1.         Unweighted (UW):

$$\omega_j = 1, \quad 1 \leq j \leq 508.$$

2.         Beta distribution type of weights (BETA):

$$\omega_j = \mathrm{dbeta}(\mathrm{MAF}_j, 1, 25)^2 \quad 1 \leq j \leq 508,$$

        where the weight follows a beta distribution with shape parameters 1 and 25.

3.         Weighted sum statistics type of weights (WSS):

$$\omega_j = \frac{1}{\mathrm{MAF}_j(1 - \mathrm{MAF}_j)} \quad 1 \leq j \leq 508.$$

4.         Logarithm of MAFs as weights (LOG):

$$\omega_j = -\log_{10}(\mathrm{MAF}_j) \quad 1 \leq j \leq 508.$$

The four weight functions can be visualized in Figure 2. From Figure 2, UW gives equal consideration to all the sequence variants, while the other weight functions (i.e., BETA, WSS, and LOG) give higher priority to low-frequency sequence variants. Among the latter three weight functions, LOG gives the most consideration to common variants with MAF greater than 5%; WSS gives nearly zero weight to variants with MAF greater than 1%; BETA gives slightly more weight than LOG for MAF less than about 1.5%, thereafter decreases much faster than LOG, and approaches zero for variants with MAF greater than 10%. These weight functions give distinct consideration to the sequence variants. It is apparent that the statistical power on adopting various weight functions would be affected by the underlying disease models. To compare the power of two methods, we randomly selected fifty sequence variants as causal variants, and simulated their phenotypes according to the following models:

a.         For quantitative phenotypes:

$$y_i = \mu + \sum_{k=1}^{50} \beta_{j_k} g_{i,j_k} + \varepsilon_i;$$

        where $e_i \sim N(0, 1)$.

b.         For binary phenotypes:

$$\text{logit}(p(y_i=1)) = \mu + \sum_{k=1}^{50} \beta_{j_k} g_{i,j_k},$$

where $g_{i,j_k}$ denotes the genotype of the $k$-th causal variant for subject $i$ and was coded as additive (i.e., $g_{i,j_k}=0$ for AA, $g_{i,j_k}=1$ for Aa, and $g_{i,j_k}=2$ for aa). $\beta_{j_k}$ was used to measure the effect of the $k$-th causal variants, $1 \leq k \leq 50$. For quantitative phenotypes, the intercept $\mu$ was set to be 0. For binary phenotypes, $\mu$ was adjusted to ensure that the case/control ratio was approximately 1:2.

In our simulations, we considered four disease models by varying the effect sizes, $\beta_{j_k}$. Each disease model was included to favor one of the weight functions.

S1      The effect sizes of all causal variants were equal:

$$\begin{aligned}\beta_{j_k} &= \beta_{S_1}, \quad \text{for} \quad 1 \leq k \leq 50; \quad \text{and}\\ \beta_j &= 0, \quad \text{if} \quad j \notin \{j_1, j_2, \ldots, j_{50}\}.\end{aligned}$$

S2      The effect sizes of causal variants were proportional to the BETA weights described above:

$$\begin{aligned}\beta_{j_k} &= \beta_{S_2} \times \text{dbeta}(\text{MAF}_{j_k}, 1, 25)^2, \quad \text{for} \quad 1 \leq k \leq 50; \quad \text{and}\\ \beta_j &= 0, \quad \text{if} \quad j \notin \{j_1, j_2, \ldots, j_{50}\}.\end{aligned}$$

S3      The effect sizes of causal variants were proportional to the WSS weights described above:

$$\begin{aligned}\beta_{j_k} &= \frac{\beta_{S_3}}{\text{MAF}_{j_k}(1-\text{MAF}_{j_k})}, \quad \text{for} \quad 1 \leq k \leq 50; \quad \text{and}\\ \beta_j &= 0, \quad \text{if} \quad j \notin \{j_1, j_2, \ldots, j_{50}\}.\end{aligned}$$

S4      The effect sizes of causal variants were proportional to the LOG weights described above:

$$\begin{aligned}\beta_{j_k} &= \beta_{S_4} \log_{10}(\text{MAF}_{j_k}), \quad \text{for} \quad 1 \leq k \leq 50; \quad \text{and}\\ \beta_j &= 0, \quad \text{if} \quad j \notin \{j_1, j_2, \ldots, j_{50}\}.\end{aligned}$$

In S1–S4, the parameters $\beta_{S_1} \ldots \beta_{S_4}$ were adjusted to ensure the power of the two methods were within a reasonable range. In S1, because all causal variants had equal effect size, common variants with higher MAFs were expected to contribute more to the phenotypes than rare variants. On the other hand, rare variants played more important roles than common variants in S2–S4 due to their relatively larger effect sizes.

To evaluate type I error, the phenotypes were simulated independently of the genetic variants. The quantitative phenotypes were simulated as, $y_i = e_i$, while the binary phenotypes were simulated as $p(y_i = 1) = 1/3$. Both GGRF and SKAT were then applied to 1,000 replicates of the simulated data to evaluate type I error.

**Simulation II: Varying Causal Variants/Noise Variants Ratios**—In practice, the investigator usually does not know how many sequence variants are associated with the phenotypes. Both GGRF and SKAT are able to handle multiple sequence variants. However, including noise variants may affect the performance of both methods. In simulation II, we evaluated the performance of the two methods by gradually increasing the number of noise variants in the analysis. Similar to simulation I, four disease scenarios (i.e., S1–S4) were simulated by randomly selecting 50 variants as the causal variants. For each disease scenario, we started with the analysis without any noise variants, and then gradually increased the number of noise variants to 50, 150, 250, 350, and 458. The corresponding casual/noise variants ratios are approximately 1:0, 1:1, 1:3, 1:5, 1:7, and 1:9. For both methods, the BETA weights (i.e., the default option of SKAT) were used to evaluate power and type I error.

**Simulation III: Various Similarity Metrics**—Both GGRF and SKAT can use a wide variety of similarity metrics to enhance their performance. In simulation III, we evaluated the performance of GGRF and SKAT with the use of different similarity metrics. For GGRF, we proposed a general $p$-norm distance-based similarity metric, and evaluated up to order of 4 (i.e., $P = 4$), denoted as D1S, D2S, D3S, and D4S, in the simulation. For SKAT, the IBS- and linear-kernels were evaluated. Note that the 1st order NDS (D1S) had the same form as the IBS-kernel. In each simulation, the phenotypes were simulated according to the disease model S4 as described above. While applying both methods, the BETA weights (i.e., the default option of SKAT) were used.

## Simulation Results

**Simulations I: Various Prespecified Weights Under Various Disease Models**—The simulation results for both quantitative and binary phenotypes were summarized in Figure 3. For all prespecified weights (i.e., UW, BETA, WSS, and LOG), the type I error of GGRF was well controlled at the 5% level. On the other hand, when the prespecified weights were not UW, SKAT showed conservative type I error (i.e., considerably less than 5%), especially when the weight function was chosen in favor of rare variants (e.g., WSS). We also found that when WSS was used, the type I error of SKAT could be even more conservative under binary phenotypes.

In terms of statistical power, both methods attained highest power when the chosen weights reflected the underlying disease scenarios (i.e., UW for S1, BETA for S2, WSS for S3, and LOG for S4). SKAT attained higher power than GGRF in disease scenario S1, while GGRF attained higher power than SKAT in all the other disease scenarios (i.e., S2, S3, and S4), across all weights. This indicated that SKAT outperformed GGRF when common variants contributed more to the disease than rare variants, while GGRF outperformed SKAT when the rare variants played more important roles than common variants.

**Simulation II: Varying Causal Variants/Noise Variants Ratios**—The simulation results for various causal variants/noise variants ratios were summarized in Figure 4. The results were similar for quantitative and binary phenotypes. In all simulations, the type I error of GGRF was well controlled. Similar to Simulation I, SKAT showed conservative type I error, which was consistently less than 5%. We also found the type I error of SKAT was close to 5% if no noise loci were included.

The power of both methods decreased when the number of noise variants increased. For disease scenarios S2–S4, the power of GGRF was consistently higher than that of SKAT. This result concurred with Simulation I, indicating that GGRF outperformed SKAT when the rare variants had a major contribution to the phenotype. In addition, when common variants had more influence on the disease phenotype than rare variants (i.e., S1) and the majority of variants were causal, GGRF could still have a better performance than SKAT. However, as the number of noise variants increased, the power of SKAT decreased more slowly than GGRF. When a large number of noise variants were included, SKAT would attain higher power than GGRF under disease scenario S1.

**Simulation III: Various Similarity Metrics**—The simulation results for various similarity metrics were summarized in Figure 5. From simulations I and II, we observed the conservative type I error of SKAT. In a recent extension of SKAT, a bootstrap approach was proposed to address the issue of conservative type I error [Lee et al., 2012]. In this simulation, we also evaluated the performance of SKAT after bootstrap adjustment. Note that bootstrap adjustment was only available for SKAT with the binary phenotype, the linear-kernel, and the BETA weight. A large number of bootstrap samples would also significantly increase the computational effort.

Similar to Simulations I and II, GGRF had well-controlled type I error (~5%) for all similarity metrics (i.e., IBS, D2S, D3S, and D4S). On the other hand, SKAT had conservative type I errors for both IBS- and linear-kernels. However, after the bootstrap adjustment, the type I error of SKAT was corrected to ~5%.

Although GGRF can use any order of NDS, its performance depends on how well the chosen NDS reflects the underlying genetic similarity. When the order increases, the model tends to put higher weights on the remote individuals (i.e., individuals who are genetically less similar). The model reaches its optimal performance when the estimated genetic similarity (i.e., weights) approaches the underlying genetic similarity. In our simulation that assumed an additive model, GGRF attained the highest power when D2S was used. The power of GGRF was comparable on using either IBS or D2S, both of which attained substantially higher power than D3S or D4S. Therefore, a low order of NDS, such as IBS or D2S, is preferred for the practical use when the mode of inheritance is additive. On the other hand, without bootstrap adjustment the power of SKAT was comparable on using either the IBS-or linear-kernel. The bootstrap adjustment of SKAT with the linear-kernel would also increase its statistical power.

It should be noted that the similarity metrics used for GGRF and the kernel function of SKAT are fundamentally different. The similarity metrics for GGRF are based on distance metrics. The linear-kernel is not a distance-based metric, and should not be used for GGRF. A comparison of the two methods using various similarity metrics/kernel functions is not straightforward. Both D2S of GGRF and the linear-kernel of SKAT can be calculated efficiently, and thus are suitable for analyzing large-scale datasets. Our simulation demonstrated that without bootstrap adjustment D2S of GGRF performed better than the linear-kernel of SKAT. After bootstrap adjustment, SKAT attained a power comparable to that of GGRF, but with an increased computational requirement.

## Application to the DHS

We applied both GGRF and SKAT to a sequencing dataset from the DHS [Romeo et al., 2009]. The dataset comprised 2,658 individuals and four candidate genes, *ANGPTL3*, *ANGPTL4*, *ANGPTL5*, and *ANGPTL6*. The four genes are members of the *ANGPTL* family that has been previously suggested to play a key role in serum triglyceride (TG) metabolism in humans [Kathiresan et al., 2008; Koster et al., 2005; Romeo et al., 2009; Shimizugawa et al., 2002]. We conducted a gene-based association analysis by using both GGRF and SKAT with consideration of potential confounding effects from race and gender. A linear-kernel was used for SKAT (default option in SKAT), while a D2S metric was used for GGRF. We started the analysis with the original scale of serum TG and tested the association of each gene with the quantitative values of serum TG. To illustrate the application to the binary phenotype, serum TG was dichotomized at the highest quartile of each of the six sex-ethnicity groups. We tested each gene by using two different strategies. First, all available variants in a gene were tested for their joint association with the phenotype. Second, only nonsynonymous (NS) variants were tested. The MAF distribution of variants in each gene was given in Figure 6. For the sake of illustration, we plotted the genetic variants with MAFs greater than 0.05 as 0.05.

The results were summarized in Table 1, where the association findings reaching the nominal significance level of 0.05 were highlighted. The most significant association finding came from *ANGPTL4*, where both methods identified an association between the nonsynonymous variants of *ANGPTL4* and the TG phenotype. The association results from GGRF attained a higher significance level than those of SKAT, for both quantitative and binary phenotypes. For instance, for the binary phenotype, GGRF obtained a *P*-value of 0.001, while SKAT had a *P*-value of 0.015. When all variants (i.e., both synonymous and nonsynonymous variants) were considered, the association could only be found by SKAT, which was also associated with a less significant *P*-value (i.e., *P*-value = 0.02). A possible explanation is that the association was mainly driven by nonsynonymous variants, while the majority of the synonymous variants were not disease related.

Neither method found evidence of association for *ANGPTL3* with TG when considering both synonymous and nonsynonymous variants together. By testing only the nonsynonymous variants of *ANGPTL3*, however, GGRF was able to detect a significant association for both the quantitative TG phenotype (*P*-value = 0.037) and the binary phenotype (*P*-value = 0.03). As compared to GGRF, SKAT did not find the association of nonsynonymous variants with the quantitative TG phenotype, but did identify an association for the binary TG phenotype (*P*-value = 0.016). Similar to *ANGPTL4*, the association was mainly seen in the nonsynonymous variants of *ANGPTL*3. Nevertheless, the association of the nonsynonymous variants with *ANGPTL*3 was not as strong as that with *ANGPTL*4. GGRF also identified a marginal association for *ANGPL*5 with the binary TG phenotype when considering both synonymous and nonsynonymous variants together ($P$-value = 0.05); while SKAT did not find any association between *ANGPL*5 and either quantitative or binary TG phenotype. No association was identified for *ANGPTL*6 using either method. This might indicate that *ANGPL*6 makes no contribution to the variation of TG. It could also be possible that the majority of the variants in *ANGPTL*6 were noncausal, so that neither method had sufficient power to detect any association.

## Discussion

A random field is a stochastic process that takes values in a Euclidean space with specific geometric structure [Adler and Taylor, 2007]. It has been extensively studied in theory and has been widely used in areas such as spatial analysis and imaging analysis. However, despite its natural advantage for high-dimensional data analyses, it has rarely been adopted in genetic research. In this article, we have proposed a GGRF method for association analysis of sequencing data underlying various types of phenotypes (e.g., binary and continuous phenotypes). GGRF is built on a Euclidean space with a flexible dimensionality determined by the number of sequence variants. With such a geometric structure, the genetic similarities between subjects can be naturally measured by their $p$-norm distance, and then connected to their phenotypic similarities. As a similarity-based method, it follows the same assumptions of SKAT and SIMreg in that, if there is a gene–phenotype association, the genetic similarity between subjects contributes to their phenotypic similarity. The proposed GGRF has several appealing features over burden tests, such as avoiding the selection of thresholds for rare variants, allowing for multiple variants with different directions and magnitude of effects, and being computationally efficient without requiring a permutation test. Empirically, we have demonstrated that GGRF attained higher power than SKAT when rare variants had relatively larger effect sizes than common variants, or when the majority of variants were causal. Previous studies have shown that if the phenotype was binary and the sample size was small (e.g., less than 2,000), SKAT could yield conservative results, leading to incorrect type I error and power loss [Lee et al., 2012; Lin and Tang, 2011; Wu et al., 2011]. Our simulation results were consistent with this previous finding. In addition, we also found a conservative type I error of SKAT for quantitative phenotypes with a relatively small sample size ($n = 697$) and extremely rare variants (e.g., MAF = 0.07%). On the other hand, because of its asymptotic property, GGRF has well-controlled type I error, even for a small sample size. This feature makes GGRF more suitable for sequencing studies with a small to moderate sample size (i.e., $n < 2,000$).

GGRF has also a close connection to SKAT. SKAT is a kernel machine-based method, which models genetic effects as a variance component in a linear mixed model. It uses a kernel function to summarize the similarity between pairs of individuals. Similarly, GGRF models the covariates parametrically and genetic effects in a nonparametric fashion. When phenotypes follow a normal distribution, we can show that there is a close connection between GGRF and SKAT. We denote the kernel matrix in SKAT as $K$ and the similarity matrix in GGRF as $S$, where $K$ and $S$ are the same except the diagonal elements of S are zeros. The SKAT model can be written as a linear-mixed model,

$$Y \sim X\beta + \upsilon, \quad \upsilon \sim N(0, \sigma^2 I + \tau^2 K).$$

$I$ is an $N \times N$ identity matrix and $\sigma^2$ is the variance of $Y_i$ under the null hypothesis; $\tau^2 K$ represents the variance component for the genetic effects [Kwee et al., 2008; Liu et al., 2007; Wu et al., 2010]. By using the factorization theorem of Besag [Besag, 1974], the GGRF model can be expressed as,

$$Y \sim X\beta + \nu, \quad \nu \sim N(0, \sigma^2(I - \gamma S)^{-1}).$$

The coefficient $\gamma$ measures the magnitude of genetic effects. Under the null hypothesis of no association (i.e., $\tau = 0$ in SKAT or $\gamma = 0$ in GGRF), the two models are equivalent. Although both SKAT and GGRF evaluate the association by testing a coefficient in the covariance matrix, the two methods model genetic effects differently, resulting in different performance. In addition, SKAT adopts a score test while GGRF uses a Wald type of statistic, which also leads to different power in testing the genetic association. Another difference between SKAT and GGRF is different similarity metrics they used. Because of the geometric structure of GGRF, the similarity metric is defined based on the $p$-norm distance between genetic vectors, and thus some kernel functions (e.g., the linear kernel) of SKAT cannot be used as similarity metrics for GGRF. In fact, the linear kernel of SKAT can be viewed as the inner product of two genetic vectors. In a Hilbert Space (e.g., a genetic space), the corresponding distance metric of the linear kernel would be the commonly used Euclidean distance (i.e., 2-norm distance) as in GGRF. GGRF is not able to use a linear kernel directly, but the corresponding D2S metric, which has a same computational speed as the linear kernel, can be used.

In the analysis of sequencing data, a wide variety of weight functions have been proposed to adjust for the contributions of rare variants to diseases. It is worthwhile to study how different weights influence association tests. In this study, our simulation results have demonstrated that the performance of a weight function is inherently determined by the underlying disease model. The ideal choice of weights for sequence variants should be proportional to their effects on the phenotype. For example, ideally, a noise variant should be given a weight of zero. Previous study also indicated that weights based on estimated effects of variants help to minimize power loss due to the inclusion of noncausal variation [Liu et al., 2013]. However, the estimation of adaptive weights from the same study utilizes phenotype information, and thus the permutation test is required to account for inflated type I error. The asymptotic test is only valid when empirical weights are estimated from other independent studies. In practice, the weight function should be carefully selected based on the purpose of the study and available prior knowledge. For example, if the aim of a study is to test the joint association of both common and rare variants, the BETA and LOG weights appear to be better choices than the WSS weight, while the WSS weight might be more suitable to detect rare variants with large effects.

In this article, we studied two types of phenotypes, quantitative phenotypes and binary phenotypes, using a GEE framework. Based on GEE, our method can also be easily applied to phenotypes with various distributions (e.g., a Poisson distribution) by simply changing the link function in equation (2). Another advantage of using GEE is that it allows us in the future to further extend the GGRF to handle multiple phenotypes or repeated measurements.

In the empirical study of the DHS, both methods found evidence of association between nonsynonymous variants in the genes *ANGPTL3*, *ANGPTL4*, and TG. In this study, GGRF showed some advantage over SKAT for detecting the most significant association, i.e., the association between *ANGPTL4* and TG. Nevertheless, the performance of the two methods is essentially influenced by the underlying disease model. So far, we still have limited knowledge of *ANGPTL3* and *ANGPTL4* in regard to their role in TG and the contribution of rare variants in *ANGPTL3* and *ANGPTL4* to TG. Further studies will be needed to validate this preliminary finding and further evaluate the role of rare variants in TG.

## Acknowledgments

We appreciate critical comments from two anonymous reviewers, and would like to thank Dajiang Liu and Jonathan Cohen for helping us access the Dallas Heart Study dataset. This work was supported by startup funds from the University of Arkansas for Medical Sciences (UAMS), by the National Institute on Drug Abuse under Award Number K01DA033346, by the National Institute of Dental & Craniofacial Research under Award Number R03DE022379, and the National Research Foundation of Korea Grant funded by the Korean Government (NRF-2011-220-C00004).

## References

Adler, RJ.; Taylor, JE. Random Fields and Geometry. Springer; New York: 2007.

Almasy L, Dyer TD, Peralta JM, Kent JW Jr, Charlesworth JC, Curran JE, Blangero J. Genetic Analysis Workshop 17 mini-exome simulation. BMC Proc. 2011; 5(Suppl 9):S2.

Ansorge WJ. Next-generation DNA sequencing techniques. N Biotechnol. 2009; 25(4):195–203. [PubMed: 19429539]

Besag J. Spatial interaction and statistical analysis of lattice systems. J R Stat Soc B. 1974; 48:259–302.

Bodmer W, Bonilla C. Common and rare variants in multifactorial susceptibility to common diseases. Nat Genet. 2008; 40(6):695–701. [PubMed: 18509313]

Davies R. The distribution of a linear combination of Chi-square random variables. Appl Stat. 1980; 29:323–333.

Eichler EE, Flint J, Gibson G, Kong A, Leal SM, Moore JH, Nadeau JH. Missing heritability and strategies for finding the underlying causes of complex disease. Nat Rev Genet. 2010; 11(6):446–50. [PubMed: 20479774]

Han F, Pan W. A data-adaptive sum test for disease association with multiple common or rare variants. Hum Hered. 2010; 70(1):42–54. [PubMed: 20413981]

He, Z.; Zhang, M.; Zhan, X.; Lu, Q. Modeling and Testing for Joint Association Using a Genetic Random Field Model. 2013. http://arxiv-web3.library.cornell.edu/abs/1302.5493eprint arXiv: 1302.5493

Hindorff LA, Sethupathy P, Junkins HA, Ramos EM, Mehta JP, Collins FS, Manolio TA. Potential etiologic and functional implications of genome-wide association loci for human diseases and traits. Proc Natl Acad Sci USA. 2009; 106(23):9362–9367. [PubMed: 19474294]

Kathiresan S, Melander O, Guiducci C, Surti A, Burtt NP, Rieder MJ, Cooper GM, Roos C, Voight BF, Havulinna AS, et al. Six new loci associated with blood low-density lipoprotein cholesterol, high-density lipoprotein cholesterol or triglycerides in humans. Nat Genet. 2008; 40(2):189–197. [PubMed: 18193044]

Koster A, Chao YB, Mosior M, Ford A, Gonzalez-DeWhitt PA, Hale JE, Li D, Qiu Y, Fraser CC, Yang DD, et al. Transgenic angiopoietin-like (angptl)4 overexpression and targeted disruption of angptl4 and angptl3: regulation of triglyceride metabolism. Endocrinology. 2005; 146(11):4943–4950. [PubMed: 16081640]

Kwee LC, Liu D, Lin X, Ghosh D, Epstein MP. A powerful and flexible multilocus association test for quantitative traits. Am J Hum Genet. 2008; 82(2):386–397. [PubMed: 18252219]

Lee S, Emond MJ, Bamshad MJ, Barnes KC, Rieder MJ, Nickerson DA, Christiani DC, Wurfel MM, Lin X. Team NGESP-ELP. Optimal unified approach for rare-variant association testing with application to small-sample case-control whole-exome sequencing studies. Am J Hum Genet. 2012; 91(2):224–237. [PubMed: 22863193]

Li B, Leal SM. Methods for detecting associations with rare variants for common diseases: application to analysis of sequence data. Am J Hum Genet. 2008; 83(3):311–321. [PubMed: 18691683]

Lin DY, Tang ZZ. A general framework for detecting disease associations with rare variants in sequencing studies. Am J Hum Genet. 2011; 89(3):354–367. [PubMed: 21885029]

Liu D, Lin X, Ghosh D. Semiparametric regression of multidimensional genetic pathway data: least-squares kernel machines and linear mixed models. Biometrics. 2007; 63(4):1079–1088. [PubMed: 18078480]

Liu K, Fast S, Zawistowski M, Tintle NL. A geometric framework for evaluating rare variant tests of association. Genet Epidemiol. 2013; 37(4):345–357. [PubMed: 23526307]

Madsen BE, Browning SR. A groupwise association test for rare mutations using a weighted sum statistic. PLoS Genet. 2009; 5(2):e1000384. [PubMed: 19214210]

Manolio TA, Collins FS, Cox NJ, Goldstein DB, Hindorff LA, Hunter DJ, McCarthy MI, Ramos EM, Cardon LR, Chakravarti A, et al. Finding the missing heritability of complex diseases. Nature. 2009; 461(7265):747–753. [PubMed: 19812666]

McClellan J, King MC. Genetic heterogeneity in human disease. Cell. 2010; 141(2):210–217. [PubMed: 20403315]

Morris AP, Zeggini E. An evaluation of statistical approaches to rare variant analysis in genetic association studies. Genet Epidemiol. 2010; 34(2):188–193. [PubMed: 19810025]

Price AL, Kryukov GV, de Bakker PI, Purcell SM, Staples J, Wei LJ, Sunyaev SR. Pooled association tests for rare variants in exon-resequencing studies. Am J Hum Genet. 2010; 86(6):832–838. [PubMed: 20471002]

Romeo S, Yin W, Kozlitina J, Pennacchio LA, Boerwinkle E, Hobbs HH, Cohen JC. Rare loss-of-function mutations in ANGPTL family members contribute to plasma triglyceride levels in humans. J Clin Invest. 2009; 119(1):70–79. [PubMed: 19075393]

Schork NJ, Murray SS, Frazer KA, Topol EJ. Common vs. rare allele hypotheses for complex diseases. Curr Opin Genet Dev. 2009; 19(3):212–219. [PubMed: 19481926]

Schuster SC. Next-generation sequencing transforms today's biology. Nat Methods. 2008; 5(1):16–18. [PubMed: 18165802]

Shimizugawa T, Ono M, Shimamura M, Yoshida K, Ando Y, Koishi R, Ueda K, Inaba T, Minekura H, Kohama T, et al. ANGPTL3 decreases very low density lipoprotein triglyceride clearance by inhibition of lipoprotein lipase. J Biol Chem. 2002; 277(37):33742–33748. [PubMed: 12097324]

Tzeng JY, Zhang D, Chang SM, Thomas DC, Davidian M. Gene-trait similarity regression for multimarker-based association analysis. Biometrics. 2009; 65(3):822–832. [PubMed: 19210740]

Tzeng JY, Zhang D, Pongpanich M, Smith C, McCarthy MI, Sale MM, Worrall BB, Hsu FC, Thomas DC, Sullivan PF. Studying gene and gene-environment effects of uncommon and common variants on continuous traits: a marker-set approach using gene-trait similarity regression. Am J Hum Genet. 2011; 89(2):277–288. [PubMed: 21835306]

Wessel J, Schork NJ. Generalized genomic distance-based regression methodology for multilocus association analysis. Am J Hum Genet. 2006; 79(5):792–806. [PubMed: 17033957]

Wu MC, Kraft P, Epstein MP, Taylor DM, Chanock SJ, Hunter DJ, Lin X. Powerful SNP-set analysis for case-control genome-wide association studies. Am J Hum Genet. 2010; 86(6):929–942. [PubMed: 20560208]

Wu MC, Lee S, Cai T, Li Y, Boehnke M, Lin X. Rare-variant association testing for sequencing data with the sequence kernel association test. Am J Hum Genet. 2011; 89(1):82–93. [PubMed: 21737059]

Zawistowski M, Gopalakrishnan S, Ding J, Li Y, Grimm S, Zollner S. Extending rare-variant testing strategies: analysis of noncoding sequence and imputed genotypes. Am J Hum Genet. 2010; 87(5): 604–617. [PubMed: 21070896]

## Appendix A. The Influence of Different Similarity Metrics on the Performance of GGRF

In GGRF, individuals are mapped into a genetic space spanned by their sequenced genotypes. The conditional phenotype mean of each individual can then be modeled as a linear function of a weighted sum of phenotypes of the remaining individuals, where the weights are determined by the genetic similarity between individuals. If there is an association between sequenced genotypes and phenotypes, we expect individuals who are adjacent to a particular individual (i.e., those are genetically similar to the individual) provide more weights than individuals who are further apart (i.e., those are genetically different to the individual). Because of the unique feature of GGRF, the performance of GGRF relies on the measurement of genetic similarity and how it reflects the underlying genetic similarity. Below we briefly discuss the choice of similarity metrics and how they influence the performance of GGRF.

In a sequencing study where each causal variant is only carried by a small number of individuals, individuals tend to distribute sparsely in the genetic space. If an individual has a limited number of nearby individuals, the model will then uses the remote individuals to approximate the individual. By using the general $p$-norm distance-based genetic similarity (NDS) proposed in the equation (1), remote individuals who carry causal variants on the other loci of the region are assigned positive weights and are contributed to the model. When rare variants in the region have the same direction of effects, an individual can be modeled not only by individuals carrying the same rare variants but also individuals carrying other causal variants. This is especially helpful for testing variants with very low frequency (e.g., singleton rare variants), where individuals carrying other rare variants and having similar phenotypes are used for improved performance. In addition, power of GGRF based on NDS increases when numbers of causal variants in the region increases. As we observed in the simulation II (Fig. 4), GGRF gained substantial power increase over SKAT when the proportion of causal variants is high.

Although NDS proposed in the equation (1) have the advantages for detecting rare variants with the same direction of effects, it is less powerful for detecting rare variants with bidirectional effects. In the later scenario, each individual is approximated by individuals carrying other causal variants and having potentially different phenotypes. This is much less an issue with the increase of MAF. With the increase of MAF, an individual carrying the causal variant is surrounded by enough number of individuals carrying the same variants and remote individuals carrying other variants play less important role in the model. Through a preliminary simulation (data is not shown), GGRF is robust to bidirectional effects when MAF reaches 0.05. To detect rare variants with bidirectional effects, we propose a centered NDS,

$$S^{cen} = (I - J)S(I - J),$$

where $S$ is an $N \times N$ similarity matrix with $s_{i,j}$ as defined in equation (1), $I$ is $N \times N$ identity matrix, and $J$ is $N \times N$ matrix with all elements being $1/N$; the diagonal elements of $S^{cen}$ were further set to 0. With the centered NDS, distant individuals with different phenotypes are assigned negative weights. As demonstrated by our simulations, GGRF using the centered NDS was robust to rare variants with bidirectional effect. Moreover, we observed that GGRF using the centered NDS was robust to noncausal variants, but did not have same advantage as NDS when majority of rare variants were causal (Fig. B1). Note that the centered NDS is also useful for identifying variants with very low frequency by incorporating the information from individuals with negative weights, i.e., the individuals not carrying the variant also contribute to the approximation but with negative weights.

We conclude that GGRF with NDS attains improved power when the effects of rare variants in a genetic region have the same direction, but is less robust to noncausal variants and bidirectional effects. GGRF with the centered NDS is robust to noncausal variants and bidirectional effects but is less powerful for detecting rare variants with same direction of effects. Ideally, we can optimally combine these two similarity metrics for maximum performance under both scenarios. We will further study this optimal GGRF and compare its performance with that of SKAT-O in our future work.

## Appendix B. Comparison of GGRF, SKAT, and Burden Test for Both One-direction and Bidirection of Effect Sizes

### Simulation Setting

To evaluate the power of three methods, we randomly selected 30 variants as causal variants and gradually increased the number of noise variants to ensure the total number of variants were 100, 150, 300, and 508. Therefore, the proportions of causal variants were approximately 30%, 20%, 10%, and 6%. Simulation studies were conducted for both one-direction and bidirection of effect sizes. For one-direction of effect size, all 30 causal variants had positive effects. For bidirection of effect sizes, 15 of 30 causal variants were randomly selected with negative effects. The phenotypes were simulated under disease model S4, and BETA weights were used for all three methods. An IBS kernel was used for SKAT, while a centered-IBS was used for GGRF.

To evaluate the type I error of three methods, the phenotypes were simulated indecently from the genotypes, and the number of variants increased from 100, 150, 300, and 508.

## Simulation Results

The simulation results for both quantitative and binary phenotypes were summarized in Figure B1. The results showed the type I errors of GGRF and Burden test were well controlled at the 5% level. On the other hand, SKAT showed conservative type I errors, especially for binary phenotypes.

In terms of statistical power, all three methods showed decreasing power as the number of noise variants increased. For one-direction of effect sizes, Burden test attained the highest power among three methods, especially when the proportional of causal variants was high (e.g., 30%). Moreover, SKAT showed higher power than GGRF for both quantitative and binary phenotypes. For bidirection of effect sizes, Burden test suffered from a significant power loss, while both GGRF and SKAT showed comparable power to their one-direction counterparts, especially when the number of noise variants was small (e.g., 100 total variants). These results indicated both GGRF and SKAT were robust to bidirection of effect sizes. Further, SKAT showed higher power than GGRF for quantitative phenotypes, while GGRF showed higher power than SKAT for binary phenotypes.

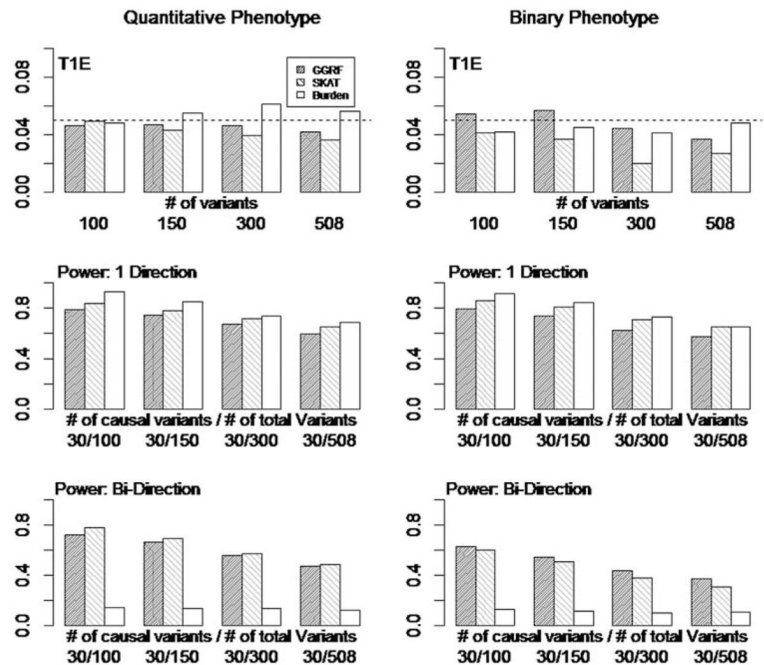

**Figure B1.**
Type I error and Power of GGRF, SKAT, and Burden test with decreasing ratio of casual variants/noise variants. Left: Quantitative Phenotypes, Right: Binary Phenotypes; T1E: Type I Error; 1 Direction: one-direction of effect sizes; Bidirection: bidirection of effect sizes.

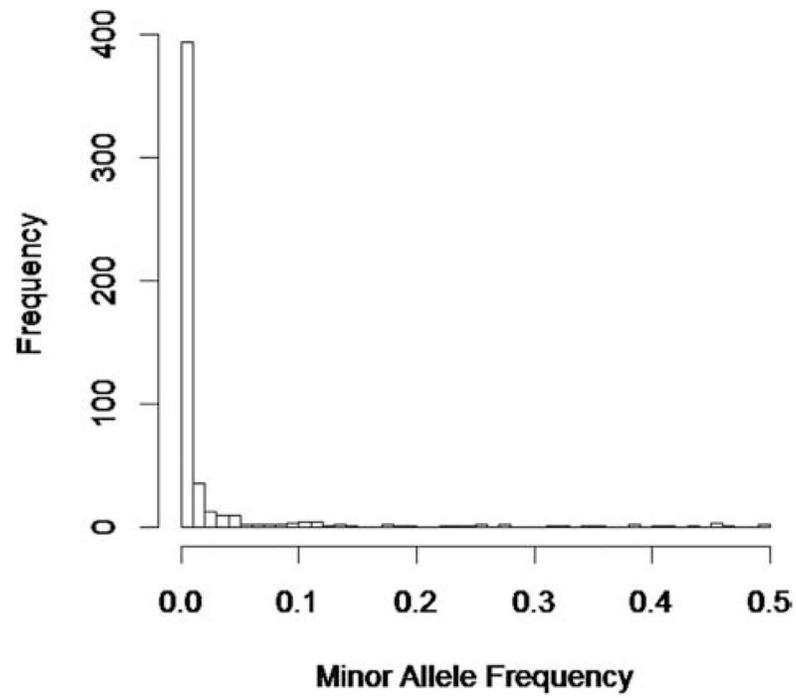

**Figure 1.**
Distribution of the minor allele frequencies of 508 sequence variants on chromosome 22 in exome sequencing data from the 1,000 Genome Project.

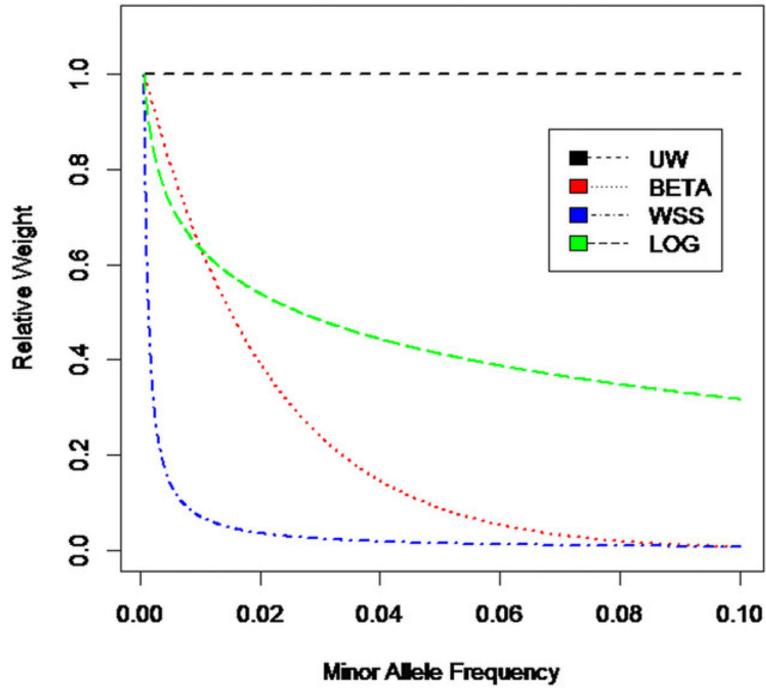

**Figure 2.**
Shape of four types of weight functions used in the simulations. Maximum weight at MAF of 0.07% was rescaled to be 1 for each weight function. The scaling does not change the relative contribution of variants.

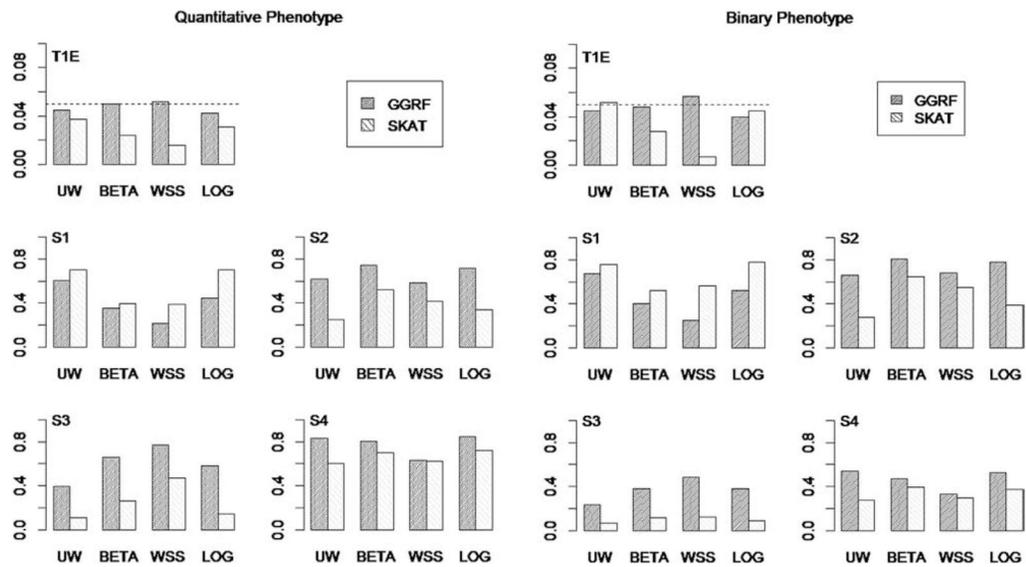

**Figure 3.**
Type I error and Power of GGRF and SKAT on using four SNP-specific weights under four disease models. Left: Quantitative Phenotypes, Right: Binary Phenotypes; T1E: Type I Error; S1–S4: power under various disease scenarios. S1: effect sizes of causal variants are all equal; S2: effect sizes of causal variants are proportional to BETA weights; S3: effect sizes of causal variants are proportional to WSS weights; S4: effect sizes of causal variants are proportional to LOG weights.

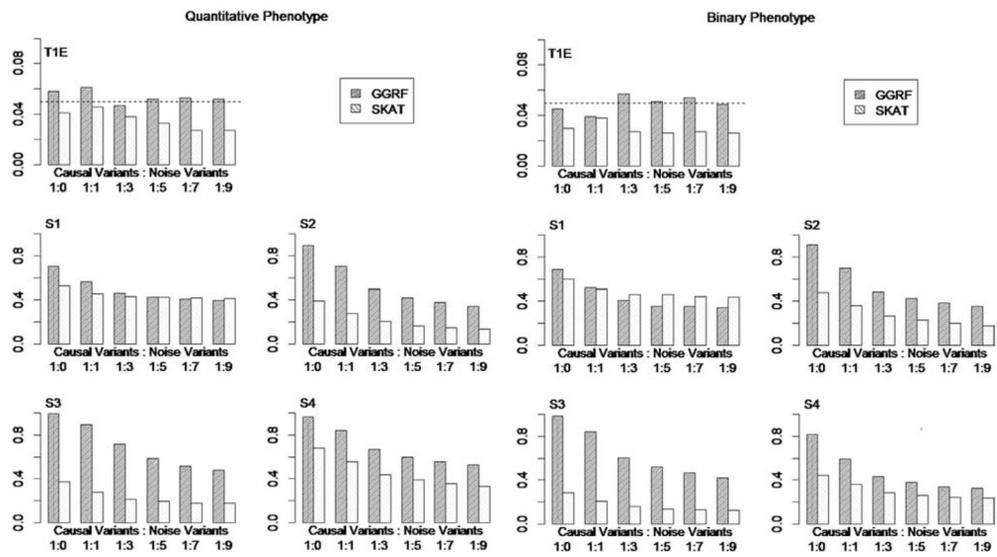

**Figure 4.**
Type I error and Power of GGRF and SKAT with decreasing ratio of casual variants/noise variants. Left: Quantitative Phenotypes, Right: Binary Phenotypes; T1E: Type I Error; S1–S4: power under various disease scenarios. S1: effect sizes of causal variants are all equal; S2: effect sizes of causal variants are proportional to BETA weights; S3: effect sizes of causal variants are proportional to WSS weights; S4: effect sizes of causal variants are proportional to LOG weights.

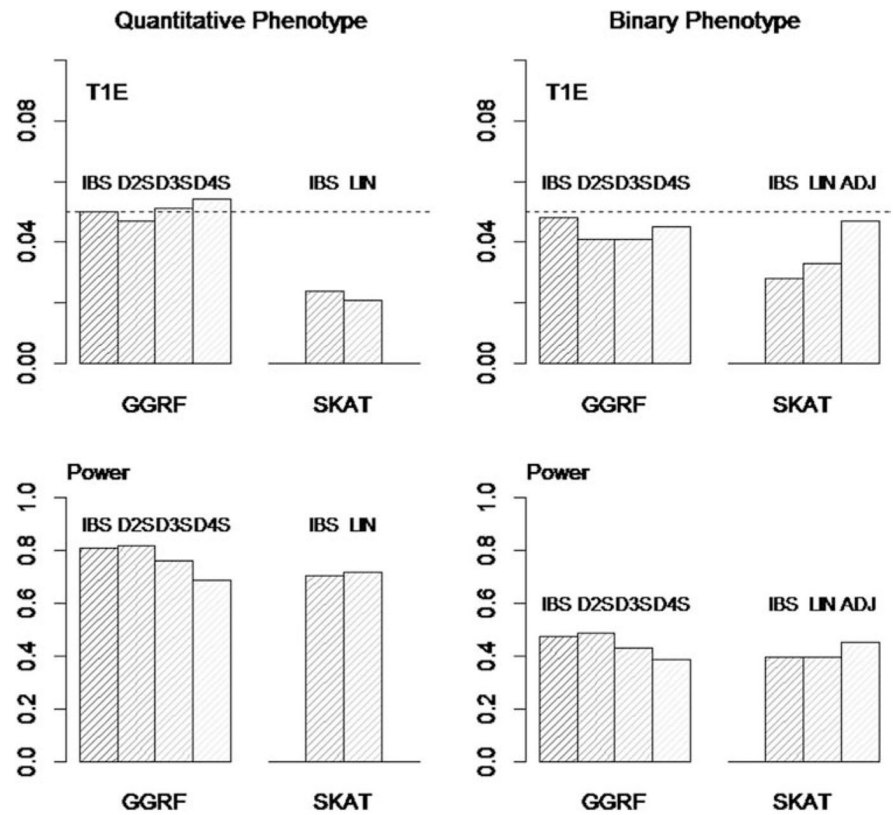

**Figure 5.**
Type I error and power of GGRF/SKAT with various similarity-metrics/kernel-metrics. Top left: type I error for quantitative phenotypes; Bottom left: power for quantitative phenotypes. Top right: type I error for binary phenotypes; Bottom right: power for binary phenotypes. ADJ: bootstrap adjustment for SKAT, only available with binary phenotypes, linear kernel, and BETA weight.

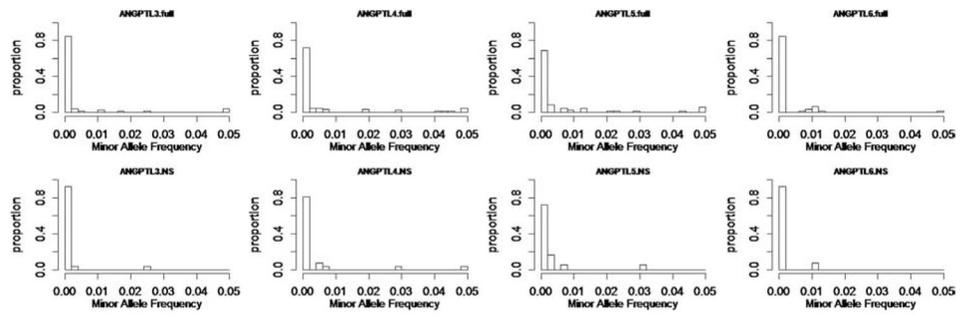

**Figure 6.**
Distribution of minor allele frequencies in *ANGPTL*3, *ANGPTL*4, *ANGPTL*5, and *ANGPTL*6 genes in 2,658 subjects from the DHS sequencing data.

**Table 1**

Application of GGRF and SKAT to the DHS sequencing data

| | # of variants | MAF range | Quantitative TG | | Binary TG | |
| | | | GGRF | SKAT | GGRF | SKAT |
|---|---|---|---|---|---|---|
| *ANGPTL3* – Full[a] | 72 | 0.019–35.8% | 0.214 | 0.120 | 0.622 | 0.061 |
| *ANGPTL3* – NS[b] | 28 | 0.019–2.46% | **0.037** | 0.103 | **0.030** | **0.016** |
| *ANGPTL4* – Full | 83 | 0.019–27.5% | 0.801 | 0.131 | 0.586 | **0.020** |
| *ANGPTL4* – NS | 27 | 0.019–27.5% | **0.008** | **0.019** | **0.001** | **0.015** |
| *ANGPTL5* – Full | 70 | 0.019–31.5% | 0.308 | 0.981 | **0.050** | 0.761 |
| *ANGPTL5* – NS | 18 | 0.019–3% | 0.193 | 0.905 | 0.101 | 0.561 |
| *ANGPTL6* – Full | 59 | 0.019–5.6% | 0.382 | 0.513 | 0.297 | 0.527 |
| *ANGPTL6* – NS | 27 | 0.019–1.04% | 0.827 | 0.156 | 0.726 | 0.116 |

[a]All genetic variants in gene *ANGPTL3*.

[b]Nonsynonymous variants in gene *ANGPTL3*.